\documentclass[runningheads]{svmult}

\usepackage{makeidx}   
\usepackage{graphicx}  
\usepackage{subeqnar}  
\usepackage{multicol}  
\usepackage{physprbb}  
\makeindex             

\begin{document}
\title*{Relativistic outflows from X-ray binaries\\
(a.k.a. `Microquasars')
}
%
%
%
%
\titlerunning{XRB jets}
%
\author{Rob Fender}
\authorrunning{Rob Fender}
%
%
\institute{
Astronomical Institute `Anton Pannekoek' and Center for High Energy Astrophysics, \\
University of Amsterdam, Kruislaan 403, 1098 SJ Amsterdam, The Netherlands
}

\maketitle              

\begin{abstract}

In this review I summarise the observational connections between
accretion and relativistic outflows -- jets -- in X-ray binaries. I
argue that jets are likely to be a fairly ubiquitous property of X-ray
binaries as a whole, an assertion which can be tested by further
observations of the Atoll-type X-ray binaries.  I discuss broad
patterns that are emerging from these observational studies, such as a
correlation between `hard' X-ray states and the presence of radio
emission, and the related anti-correlation between jet strength and
mass-accretion rate as inferred from X-ray studies alone. I briefly
discuss possible future directions for research and compare X-ray
binary jets to those from Active Galactic Nuclei and Gamma Ray Bursts.

\end{abstract}

\section{History and Introduction}

There has been a great deal of renewed interest in the past half a
decade or so in the phenomena of relativistic outflows, or `jets' from
binary systems in our own galaxy. These are often referred to as
`microquasars' because of the apparent similarities with the Quasars,
or with Active Galactic Nuclei (AGN) in general. The particular type
of stellar binary systems in which these jets seem to orginate are the
X-ray binaries (XRBs), so-called because they are powerful sources of
X-ray radiation. In XRBs a more-or-less `normal' star loses matter to
a compact collapsed companion, either a neutron star or a black hole;
it is generally accepted that the accretion of material by the compact
star, a process far more energetically efficient than nuclear fusion,
is the source of the enormous power output of these systems (which can
exceed in some cases $10^{38}$ erg s$^{-1}$, or the Eddington
luminosity for a one solar mass object). Since the power source of AGN
is similarly believed to be accretion of matter by a collapsed object,
in this case a supermassive ($10^6$--$10^9$ M$_{\odot}$) black hole,
the term `microquasar' is more than simply an indicator of similar
morphologies (ie. accretion, jet) but maybe also of similar
physics. Therefore understanding such sources is important not only in
the context of accretion physics and the evolution of `local' systems,
but maybe also for our broader understanding of the physics and
evolution of the powerhouses of the universes, the AGN.

Fig 1 is a schematic of a generic X-ray binary, indicating the
(probable) sites from which emission at different energies
originates. Note that the observable spectral extent of such systems
can be very broad -- the classical black hole candidate (BHC) XRB, Cyg
X-1, is a well-detected source from $\leq 1$ GHz in the radio band to
$\geq 1$ MeV in $\gamma$-rays, a range of $10^{12}$ in photon
energy. Most schematics, certainly until a few years ago, would not
have included the jet, and one of the goals of this paper will be to
discuss exactly how ubiquitous is this feature in XRBs. It is
interesting to note that the jet, when present, is by far the largest
structure directly associated with the XRB in general, and accretion
process in particular, and the only one of the structures indicated in
Fig 1 which has actually been directly observed (in radio images --
e.g. Fig 2).

\begin{figure}
\begin{center}
\includegraphics[width=.75\textwidth, angle=270]{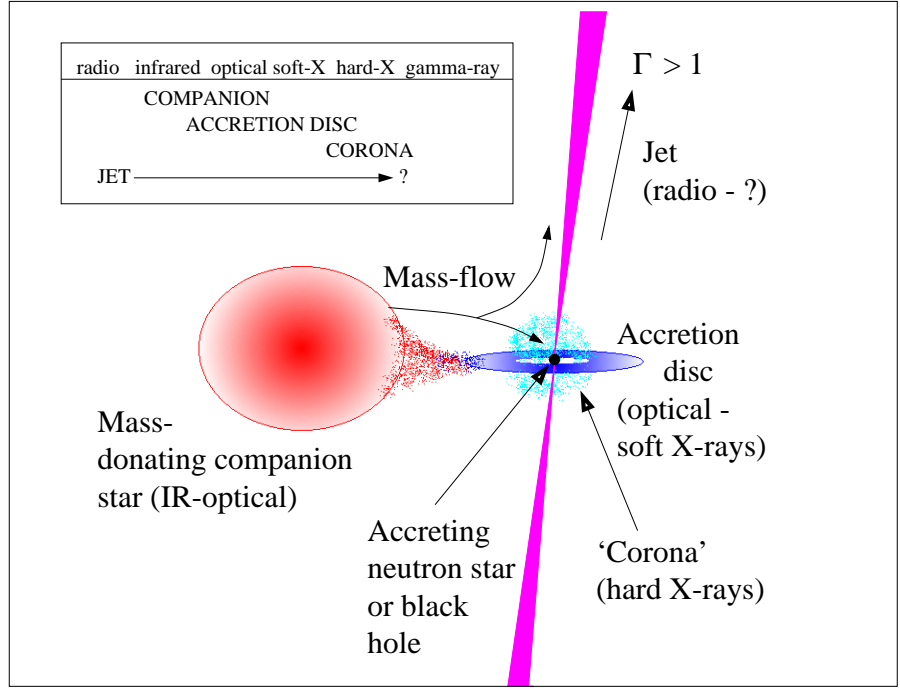}
\end{center}
\caption{
A schematic of the generally accepted structure of a `typical' X-ray 
binary system, indicating the locations of the sites believed to 
correspond to observed emission at different wavelengths.
}
\label{fig1}
\end{figure}

\begin{figure}
\begin{center}
\includegraphics[width=.8\textwidth, angle=270]{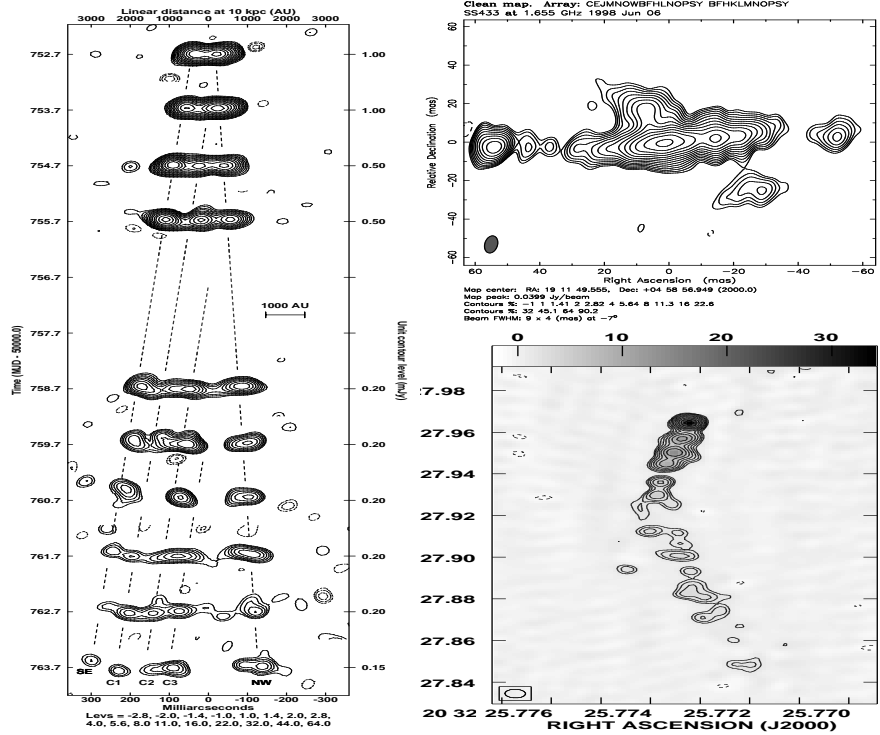}
\end{center}
\caption{Recent radio observations of three famous XRB jet sources. Left: A
sequence of images of (apparent superluminal) ejections from GRS
1915+105 observed with MERLIN (Fender et al. 1999a). Top right: A
recent combined EVN/VLBA image of SS 433 (Paragi et al. 2001). Lower
right: a VLBA image of a one-sided curved jet from Cyg X-3 following a major
radio flare (Mioduszewski et al. 2001).  }
\label{fig1}
\end{figure}

Jets from XRBs as a phenomenon were first discovered from SS 433
(Spencer 1979; Hjellming \& Johnston 1981a,b), a highly unusual system
in many ways. The source displays optical (and infrared and X-ray)
emission lines which show periodic Doppler shifts indicating a
precessing bipolar outflow with velocity $v=0.26c$; the radio jets
appear to precess as predicted from the optical lines. The
(apparently\footnote{scepticism is healthy}) rather well-defined and
only mildly relativistic velocity (bulk Lorentz factor $\Gamma = (1 -
v^2/c^2)^{-1/2} = 1.04$) are unique amongst `relativistic' jet
sources. Importantly, SS 433 is the only system, XRB or AGN, for which
atomic emission lines have been associated with the outflow, thereby
establishing a baryonic content (more of this later). The significant
and variable linear polarisation of the jets confirmed the synchrotron
interpretation for the origin of the radio emission.

Over the subsequent 15 years, a handful of other XRBs (e.g. Cyg X-3,
Cir X-1, 1E 1740.7-2942, GRS 1758-258), were identified as being
associated with radio jets. In the case of Cyg X-3, an apparent
velocity of $\sim 0.3c$ was measured (Geldzahler et al. 1983). Perhaps
all XRB jets would turn out to have a velocity of $\sim 0.3c$ ? This
picture was comprehensively refuted in 1994 when Mirabel \& Rodr\'\i
guez (1994) discovered apparent superluminal motions in multiple
ejections from the XRB GRS 1915+105. While distance-dependent,
interpretations of the intrinsic velocity of the ejecta suggested $v
\geq 0.9c$, significantly relativistic ($\Gamma \geq 2$). Clearly XRBs
could eject material at extremely high velocities, comparable to those
observed in AGN (where the phenomenon of apparent superluminal motion
is relatively commonly observed and relatively easily explained as a
geometric effect -- e.g. Rees 1966; Zensus \& Pearson 1987; Gomez et
al. 2000). Shortly after the observations of GRS 1915+105, a second
`superluminal' XRB, GRO J1655-50, was discovered (Tingay et al. 1995;
Hjellming \& Rupen 1995). Since GRO J1655-40 was demonstrated to be a
strong BHC (Bailyn et al. 1995 and several subsequent papers), it was
widely concluded that GRS 1915+105 was a BHC, something supported but
never confirmed dynamically by further observations. Less certainly,
it was asserted that the apparent dichotomy between the $\sim 0.3c$
sources (i.e. SS 433 and Cyg X-3) and the $\geq 0.9c$ sources (ie. GRS
1915+105 and GRO J1655-40) reflected the difference in escape speeds
from the vicinity of neutron stars and black holes respectively
(e.g. Livio 1999).  However, at this stage it was still generally
perceived that relativistic jets, as a property of X-ray binaries,
were a rare phenomenon, a feature common only to a small group of
`unusual' systems. Recent radio images of three of the most famous jet
sources are presented in Fig 2. 
It is worth reminding the reader that, as in the cases of AGN and
GRBs, the outflows are relativistic in {\em two} senses -- ie. they
have relativistic ($1 < \Gamma < 100$) bulk velocities (ie. the proper
motions we can resolve in radio images) and in addition are comprised
of populations of relativistic particles ($1 < \gamma < 10000$) which,
in spiralling around field lines in the magnetised plasma, produce the
observed synchrotron emission.

It now seems likely that jets from XRBs are not so rare, and that for
certain broad classes of X-ray binaries, maybe even the majority, the
jet is as integral a part of the mass transfer process as the
accretion disc. Furthermore, the apparent dichotomy in bulk velocities
also no longer appears to be valid, with Fomalont, Geldzahler \&
Bradshaw (2001a,b) clearly estalishing outflow velocities
significantly higher than $0.3c$ from the neutron star binary Sco X-1
(and besides, there is no clear evidence that either SS 433 or Cyg X-3
host neutron stars and not black holes, anyway!).  In this review I
shall try to summarise the state of existing knowledge, and what
appear to be fruitful avenues for further observational and
theoretical study. The key question before we can advance to this
stage is however, just how important are jets for the physical
processes occurring in X-ray binaries ? I hope to answer this in the
next section by establishing their near-ubiquity.

\section{The near-ubiquity of jets from X-ray binaries}

In the following, I shall make the assumption that any detection of
radio emission corresponds to evidence for jet production (in much the
same way as detection of X-rays is taken as evidence for accretion
processes). This is based upon the qualitative argument that whenever
we have resolved radio emission it has had a jet-like appearance
(except, perhaps, in the case of the unusual transient CI Cam in which
a more isotropic radio nebula seems to have formed -- Mioduszewski et
al., in prep), and on the following more quantitative argument -- for
a maximum brightness temperature of $\sim 10^{12}$K for the
synchrotron process, a flux density of $\sim 5$ mJy at 5 GHz (quite
weak) corresponds to a physical size of $\sim 10^{12}$ cm at a
distance of 5 kpc. This is an order of magnitude larger than the
typical orbital separation ($\sim 10^{11}$ cm) of a low-mass X-ray
binary, and so the simplest explanation is that such a large stucture
is maintained by an outflow.

In terms of accretion and jet production, the most useful separation
into classes of X-ray binaries is between neutron stars and black holes.
I have attempted to do this both in the following sections, and also
in table 1, in which approximate numbers of radio detections as a
fraction of total known populations is indicated. Catalogues of XRBs
with broad classifications can be found in van Paradijs (1995) and
Liu, van Paradijs \& van den Heuvel (2000,2001).

\begin{table}
\begin{center}
\begin{tabular}{|cc|}
\hline
Class & Fraction as radio sources \\
\hline
BHCs (persistent) & 4 / 4 \\
BHCs (transient) & $\sim 15 / 35$  \\
\hline
NS (Z) & 6 / 6 \\
NS (Atoll) & $\sim 5 / 100$ \\
NS (XRP) & $\sim 0 / 80$ \\
\hline
\end{tabular}
\caption{Approximate numbers of radio detections (=jet production) in
the different types of XRBs. Clearly detection of a large number of
Atoll sources holds the key to unambiguously establishing 
(or not) the ubiquity of jets in XRBs.}
\end{center}
\end{table}

Note that while the fraction stated in table 1 for the BHC transients
is $\sim 15/35$, for those transients for which {\em any} radio
observation was reported, the fraction is 15/16 (Fender \& Kuulkers
2001) -- how much this reflects non-publication of non-detections is
unclear. 

\subsection{Neutron stars}

The three broad classes of accreting neutron star, and the relation
between X-ray and radio properties, are summarised in Fig 3.

\begin{figure}
\begin{center}
\includegraphics[width=.7\textwidth, angle=0]{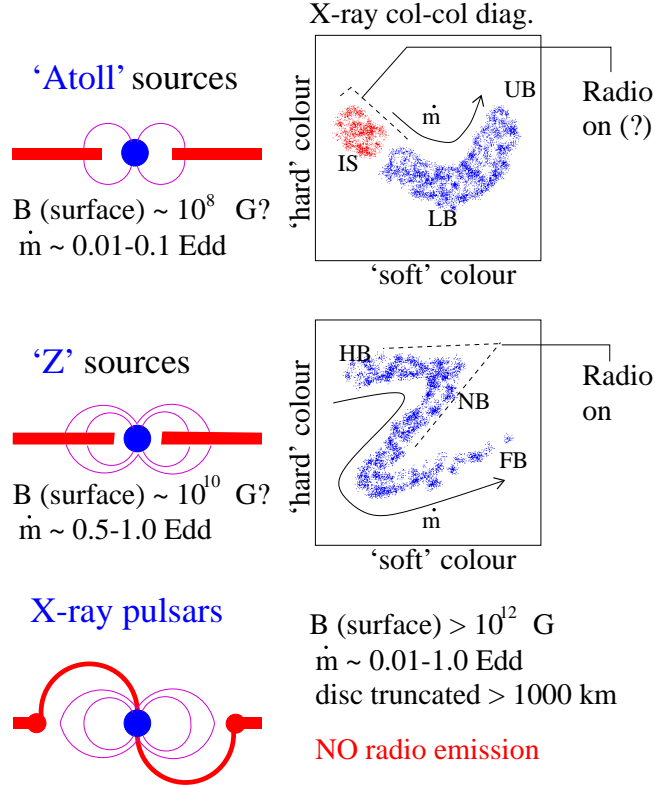}
\end{center}
\caption{ A rough schematic of the three types of neutron star X-ray
binary, a rough physical interpretation, and indication as to which
states are associated with the presence of radio emission.  For the
Atoll sources, this association is currently speculative.
Abbreviations are IS=Island State, LB=Lower Banana, UB=Upper Banana,
HB=Horizontal Branch, NB=Normal Branch, FB=Flaring Branch.  }
\label{fig1}
\end{figure}

\subsubsection{The Z sources}

These are the brightest persistent X-ray sources in the sky (the
single brightest nonsolar X-ray source, Sco X-1, is the prototype of
the group; Hasinger \& van der Klis 1989). There are six in our
galaxy, and possibly one in the LMC. The Z sources are thought to
contain neutron stars with relatively low ($\leq 10^9$ G) dipole
magnetic fields, accreting at or near the Eddington limit.  All six
galactic systems are variable but reliable radio sources, with a
comparable radio luminosity (when on the `horizontal branch' --
Penninx 1989; Fender \& Hendry 2000). The `Z' refers to the pattern
traced out in the X-ray colour-colour diagram (CD) in which three
(possibly four) branches smoothly connect. Penninx et al. (1988) found
that the radio emission was strongest on the `Horizontal Branch' and
weakest on the `Flaring Branch' in the Z source GX 17+2, an apparent
anti-correlation with accretion rate, $\dot{m}$ as deduced from X-ray
observations alone. This relation between X-ray `state' (as described
by the branches of the Z) and radio emission seems to be a property
common to all the Z sources (Hjellming \& Han 1995 and references
therein).  Recently high-resolution radio observations have revealed
unequivocal evidence for a variable jet-like structure associated with
Sco X-1 (Bradshaw, Fomalont \& Geldzahler 1999; Fomalont, Geldzahler
\& Bradshaw 2001a,b). The observations, of a variable core and moving,
variable lobes, are interpreted as the impact of a highly relativistic
beam on the ISM, producing the advance of radio `hotspots' (Fomalont
et al. 2001a,b). Given the similarities in their radio properties, and
the resolved jet in Sco X-1, the simplest conclusion (ie. using
Occam's razor) is that all Z sources produce relativistic
jets. Furthermore, the `unusual' system Cir X-1 has some Z-like
properties (Shirey, Bradt \& Levine 1999) and is a source of radio
jets from arcsecond to arcmin angular scales (Stewart et al. 1993;
Fender et al. 1998).

\subsubsection{The Atoll sources} 

In the original classification of Hasinger \& van der Klis (1989) and
subsequent works, the Atoll sources were discussed as a separate small
subgroup of bright low mass X-ray binaries with relatively low
magnetic fields, accreting at lower rates than the Z sources (e.g. van
der Klis 1995 for more details). Since then it seems likely that
Atoll-like properties may be shared by the majority of low-field
accreting neutron stars (van Paradijs, Ford, van der Klis, private
communication) and so we shall adopt this viewpoint here. If this is
the case then the Atoll sources, which will now include the groups of
`bursters', `dippers' etc., are the largest class of catalogued X-ray
binaries (see table 1). Little is known about the radio properties of
the Atoll sources, except that, as a population, they are faint
sources (typically $<1$ mJy at cm wavelengths -- Fender \& Hendry
2000). The only Atoll source to be regularly and repeatedly detected
at radio wavelengths was, until recently, the bright system GX 13+1
(Garcia et al. 1988). More recently other Atoll systems have been
discovered to have radio counterparts (e.g. 4U 1728-34/GX 354+0 --
Mart\'\i {} et al. 1998), and sources with known transient radio
counterparts have been discovered to be Atoll-like in nature (e.g. Aql
X-1 -- Reig et al. 2000). So while it is clear that as a population
Atoll sources are not particularly radio-bright, it is also clear that
they do produce detectable radio emission uncer certain
conditions. This then implies that the majority of catalogued low-mass
X-ray binaries are capable of producing a radio jet. However, until a
jet is directly resolved from an Atoll source (a key future
observation) this will remain unproven.

\subsubsection{The X-ray pulsars}

These systems possess much stronger magnetic fields ($\geq 10^{11}$ G)
than the Z or Atoll sources, which results in the disruption of the
accretion flow at a distance of several thousand km from the neutron
star surface (e.g. Bildsten et al. 1997 and references therein). As a
population they are significantly fainter than even the Atoll sources,
and no strong-field X-ray pulsar has ever been detected
as a radio synchrotron source (Fender \& Hendry 2000). Thus the strong
possibility exists that such systems do not produce jet-like outflows,
due to the extreme disruption of the accretion flow. Deep radio
observations of some nearby X-ray pulsars would be useful to further
constrain this.

\subsubsection{NS transients}

Neutron star soft X-ray transients (Chen, Shrader \& Livio 1997;
Campana et al. 1998) can probably be classified as Atoll-like
(e.g. the case of Aql X-1 -- Reig et al. 2000). As with the BHC
transients, there seems to be a discrete ejection of synchrotron
emitting material associated with the sudden increase in luminosity at
the start of the outburst. This manifests itself in a transient radio
event which becomes optically thin within a few days (presumably due
to decreasing self-absorption as the ejected component expands) and
then fades away monotonically (Hjellming \& Han 1995; Fender \&
Kuulkers 2001).

Several transients contain high field accreting X-ray pulsars
(e.g. Bildsten et al. 1997) and, as with the more persistent sources
of this type, none have ever been detected as radio synchrotron 
sources (Fender \& Hendry 2000).

\subsection{Black hole candidates}

The description of the accretion state of the BHCs differs from that
of the neutron stars as it is perceived that probably all black holes
can, under the right conditions, achieve all states -- ie. it is not
currently perceived that there are different `types' of black hole
(the only obvious distinguishing characteristic would be the black
hole spin). These `states' are summarised in Fig 4. Note that while
it was originally supposed that the states were a more or less
one-dimensional function of mass accretion rate, which could itself be
tracked via soft X-ray (disc) flux (see e.g. the pattern of behaviour
in GRO J1655-40 -- Mendez, Belloni \& van der Klis 1998), it now seems
clear that the picture is not so simple. One problem is that the same
`state' in terms of the X-ray spectral and timing properties can be
reproduced at extemely different flux levels (e.g. Homan et al. 2001).

\begin{figure}
\begin{center}
\includegraphics[width=10cm, height=12cm,angle=270]{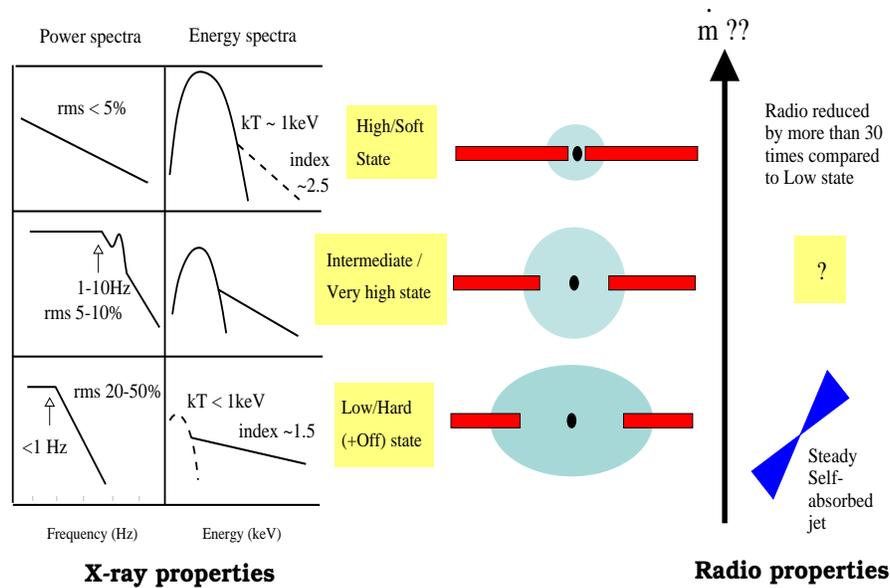}
\end{center}
\caption{
BHC spectral states, as classified by their X-ray spectral and 
timing properties. A rough physical interpretation, based on the X-ray
data alone, is indicated, as is the relation to radio emission.
BHC transients typically, although not exclusively, transit from
undetectable levels to the soft state in a short period of time.
The relation of states to mass accretion rate, previously thought to be
quite clear, is now less certain.
}
\label{fig1}
\end{figure}

The two most distinct states are the the Low/Hard and High/Soft states,
being the extremes of `nonthermal' and `thermal' spectra respectively
(this is an oversimplification). There also exists a hybrid state,
labelled the Very High or Intermediate state, which is less commonly
observed than either the Low/Hard or High/Soft states, whose relation
to radio emission, and hence presumably jet production, is unclear
(but see Corbel et al. 2001). 

\subsubsection{The Low/Hard state}

The Low/Hard X-ray state (historically called `Low' because it is
generally weaker than the High/Soft state in the soft X-ray band, and
`Hard' since it is dominated by a nonthermal power-law component which
peaks at hard X-ray ($\geq 50$ keV) energies) is the state in which
the four persistent BHCs in our galaxy spend most of their time (I
consider these four to be Cyg X-1, GX 339-4, 1E1740.7-2942 and GRS
1758-258, although it should be noted that, at the time of writing, GX
339-4 has been at extremely low levels for over a year and
consistently displays a larger amplitude of X-ray variability than the
other systems).

In the early 1970s a transition from the High/Soft 
(possibly only `Intermediate' -- see discussion in Belloni et al. 1996)
to Low/Hard X-ray
states in Cyg X-1 was observed to be coincident with the appearance of
a radio counterpart to this source (Tananbaum et al. 1972). It has
since been established that while the source is in the Low/Hard state,
which seems to be most of the time, it steadily emits a relatively low
level (typically 5-15 mJy at cm wavelengths) of radio emission
(e.g. Brocksopp et al. 1999). The spectrum of the radio emission is
remarkably flat and extends to at least the millimetre regime (Fender
et al. 2000b). Furthermore, the radio emission is
modulated at the 5.6-day orbital period of the system (Pooley, Fender
\& Brocksopp 1999). All of this evidence taken together suggests that
the flat spectrum radio emission arises in a continuously-generated,
partially self-absorbed compact jet from the system (with the orbital
modulation possibly due to variable free-free absorption in the dense
stellar wind of the OB-type mass-donor -- Brocksopp
2000). Confirmation of this hypothesis appears to have recently been
achieved with VLBA images of the system clearly resolving an
asymmetric jet from a compact core (Stirling, Garrett \& Spencer 1998;
Stirling et al. 2001).

The other three persistent Low/Hard state systems also show flat radio
spectra, and the two Galactic centre sources, 1E1740.7-2942 and GRS
1758-258, are associated with parsec-scale jet/lobe structures (the
original motivation for the name `microquasar'; Mirabel et al. 1992;
Rodr\'\i guez, Mirabel \& Mart\'\i {} 1992; Mirabel 1994). Furthermore, in both
Cyg X-1 and GX 339-4 there is an approximately linear relation between
the X-ray flux (dominated by the non-thermal power-law) and the radio
emission (Brocksopp et al. 1999; Corbel et al. 2000) indicating a
clear coupling between accretion (presumed to be reflected in the
strength and spectrum of the X-ray emission) and outflow. While GX
339-4 is the one source for which the jet has probably not been
reliably resolved, it is the one for which linear polarisation (at the
level of a few \%), has been measured (Corbel et al. 2000), supporting
the synchrotron-emitting jet model.

While most BHC transients are observed to evolve rapidly (hours) from
a `quiescent' state to the High/Soft state (and are generally
accompanied by an optically thin radio outburst -- e.g. Fender \&
Kuulkers 2001), a few X-ray transients have been observed to spend an
extended period in the Low/Hard state. A careful comparison of these
Low/Hard state transients reveals that, following an initial optically
thin radio event, they develop low-level, inverted-spectrum radio
components. These were originally dubbed `second stage' radio sources
(e.g. Hjellming \& Han 1995 and references therein). In Fender (2001)
it is argued that these components are the same as the flat/inverted
spectrum components observed from the persistent sources in the
Low/Hard state, and furthermore that such spectral components, 
and therefore compact jets, are a general property of the Low/Hard
state.

\subsubsection{The High/Soft state}

Early observations of Cyg X-1 (e.g. Tananbaum et al. 1972) suggested
that the radio emission from the source was suppressed when in the
High/Soft state (compared to the Low/Hard state). Despite the great
accessibility of this system to the world's radio telescopes, a chance
to test this hypothesis during a transition of the source to a soft
X-ray state in 1996 was missed. However, observations near the end
of the soft state support a scenario in which radio emission is
stronger during the Low/Hard state (Zhang et al. 1997).

It was a year-long transition to the High/Soft state by GX 339-4 in
1998 in which the `quenching' of the radio emission compared to the
Low/Hard state was definitively established (Fender et al. 1999;
Corbel et al. 2000).  Radio monitoring of the source revealed the cm
wavelength radio emission to have dropped by a factor of $\geq 25$
during the High/Soft state, and to return to its previous levels once
the source resumed the Low/Hard state (Fig 5).

\begin{figure}
\begin{center}
\includegraphics[width=8cm, height=12cm,angle=270]{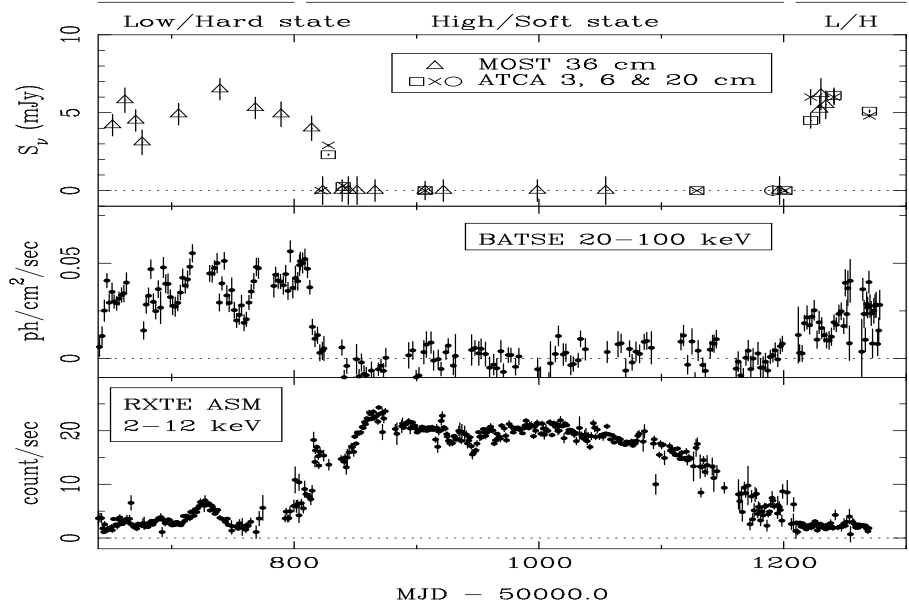}
\end{center}
\caption{ Simultaneous `quenching' of the radio (top panel) and hard
X-ray emission during a year-long high/soft state in the black hole
candidate GX 339-4 (from Fender et al. 1999; see also Corbel et
al. 2000).  }
\label{fig1}
\end{figure}

\subsubsection{The Very High/Intermediate state}

Little is clearly understood about the radio emission during the
(comparitively rare) Very High/Intermediate state of BHCs, in which
both thermal (disc) and nonthermal (power-law) spectral components can
be present. Is it the presence of the nonthermal component, or the
absence of the thermal component, which is necessary for the
production of radio emission ?  Recent observations (Corbel et
al. 2001) suggest the latter, but further observations of this state
are required.

\subsubsection{BHC transients}

BHC X-ray transients (e.g. Chen et al. 1997; Charles 1998) are generally
associated with radio outbursts (e.g. Hjellming \& Han 1995; Fender \&
Kuulkers 2001), which have been resolved on a small number of occasions
into discrete ejections, sometimes multiple, of radio emitting components
(e.g. GRO J1655-40 -- Tingay et al. 1995; Hjellming \& Rupen 1995).

In most cases the transients seem to transit from `quiescence' (which
may be some very low-level version of the Low/Hard state
described above) to the High/Soft state in the space of a few days or
less. However, in some rare cases (e.g. Fender 2001; Brocksopp et
al. 2001) a transient will `only' make it to the Low/Hard state.
Whichever `branch' the transient takes, it seems that the intial rise
is generally associated with a discrete ejection event (Fender \&
Kuulkers 2001), sometimes multiple events (e.g. Kuulkers et al. 1999).
Subsequently, if the source `achieves' the High/Soft state, there
appears to be no re-emergence of core radio emission (and so we can
assume the radio jet is `switched off' and any radio emission we
observe is physically decoupled from the ongoing accretion process);
if it instead finds itself in the Low/Hard X-ray state a flat
or inverted-spectrum component emerges (Fender 2001). 

\subsubsection{GRS 1915+105}

\begin{figure}
\begin{center}
\includegraphics[width=7.5cm, angle=270]{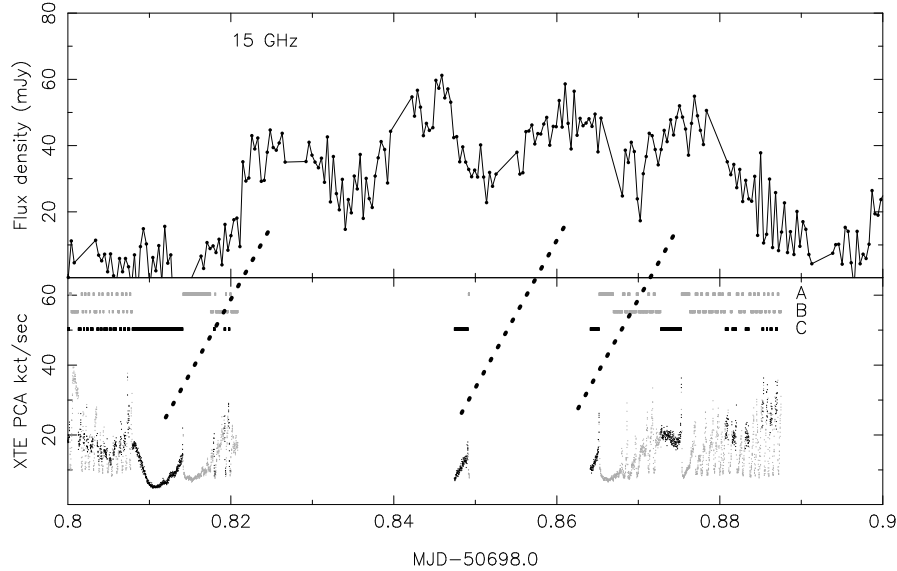}
\end{center}
\caption{
The one-to-one correspondence, in GRS 1915+105, of brief (minutes) transitions
into state `C', roughly analogous to the Low/Hard state in more
traditional BHCs, with the formation of discrete radio oscillation
events. Four radio events are observed to be associated with a dips
into the hard state -- gaps in the X-ray light curve are due to Earth
occultations, and there was almost certainly a hard dip associated
with the second radio event as well. The delay is probably due to the
time required for self-absorption of the radio emission to decrease
(as the ejecta expand).  From Klein-Wolt et al. (2001).
}
\label{fig1}
\end{figure}

While one of the aims of this review is to establish that jets are not
a bizarre property of some small subset of XRBs, but rather a more
ubiquitous feature, one source, GRS 1915+105, still deserves a mention
on its own. The source displays a remarkable and unique range of X-ray
behaviour which can however be broken down into transitions between
three broadly-defined `states' (Belloni et al. 2000).  It was also the
first system for which we had direct evidence of highly relativistic
flows (Mirabel \& Rodr\'\i guez 1994) and displays an extraordinary
variety of radio behaviour (e.g. Pooley \& Fender 1997), most of which
can (presumably) be associated with the formation of
synchrotron-emitting jets.

One of the three `states' into which any X-ray light curve of GRS
1915+105 can be deconstructed (state `C') is broadly analogous to the
Low/Hard state of traditional BHCs, being dominated by a power-law
component in the X-ray band. From a comparison of many hours of
overlapping X-ray and radio observations, we are confident that
radio oscillation events are directly, and only, associated with these
hard states in this source (Fig 6; Klein-Wolt et al. 2001), although
there are alternative opinions expressed in the literature (e.g. Naik
\& Rao 2000). 

GRS 1915+105 shows other extraordinary properties, as if attempting to
provide all the observational data we need to understand the
`disc--jet' coupling on its own. For example it was the first source
for which there was unequivocal evidence for infrared synchrotron
emission (Fender et al. 1997; Eikenberry et al. 1998, 2000; Mirabel et
al. 1998; Fender \& Pooley 1998; 2000) and is the clearest example of a
flat--spectrum core being resolved into a quasi-continuous jet
(Dhawan, Mirabel \& Rodr\'\i guez 2000; Feretti et al. 2001).

\section{Connections}

Some broad patterns are now beginning to emerge from these studies;
these patterns are clues to generic properties of jets, their coupling
to the accretion process and so on. In no particular order, these include:

\begin{itemize}
	\item{A broad correlation between hard X-ray states and radio
emission, in particular in BHCs. Meier (2001) takes such observations
as direct evidence for the MHD formation of jets in geometrically
thick accretion flows.  These hard X-ray states are generally
interpreted as arising via inverse Comptonisation (Poutanen 1998 and
references therein) in a `corona' and/or an advection-dominated
accretion flow (ADAF; e.g. Esin, McClintock \& Narayan 1997). It is
interesting to note that, to my knowledge, in every case where an ADAF
has been invoked to explain the optical--X-ray spectrum in XRBs, radio
emission is present (and yet is not fit by ADAF models).}
	\item{A related point is that in all types of X-ray binary for
which we have a clear picture,
the strength of the jet appears to be {\em anticorrelated} with the
mass accretion rate as inferred from X-ray spectral and timing studies
alone (e.g. Figs 2,3; Belloni, Migliari \& Fender 2000). 
Jets may well turn out to be an important 
factor in the state transitions associated with BHCs, Z sources and
probably also Atolls.}
	\item{The jet, however it is formed, whether via MHD or some
other `black box' really seems to carry a lot of the accretion power
-- in the case of the hard state of BHCs it seems inescapable that the
jet requires at least 10\% of the accretion energy budget -- since
current models often attempt to fit observations to a rather higher
degree of accuracy than this its effect can presumably not be ignored.}
\end{itemize}

\begin{figure}
\begin{center}
\includegraphics[width=12cm, angle=0]{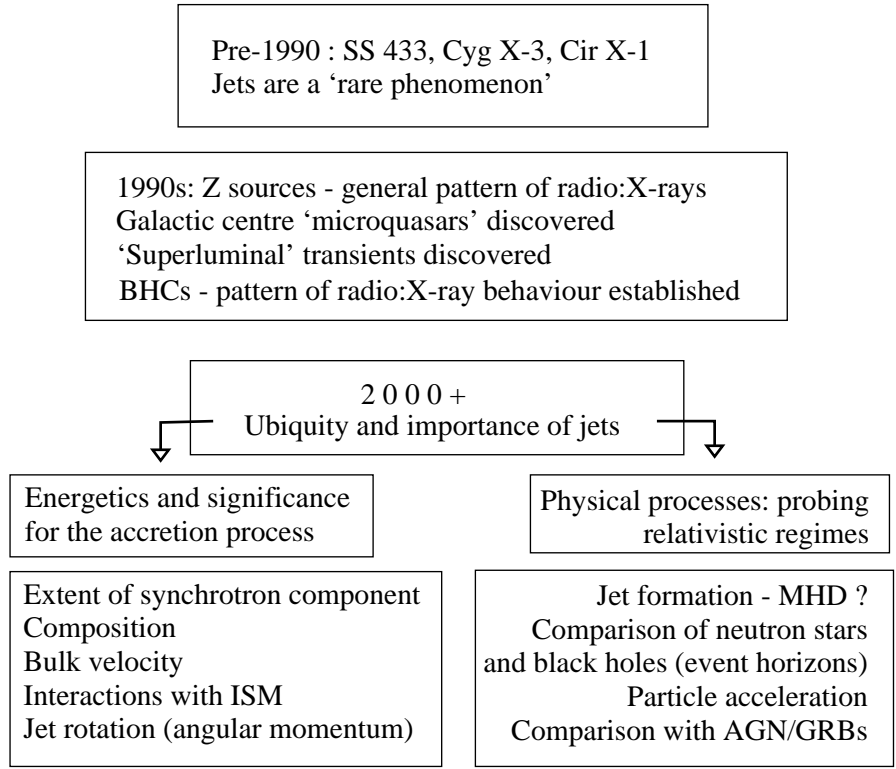}
\end{center}
\caption{
A schematic indication of where we are now with research into jets from
X-ray binaries, and some possible future directions. Central to this scheme
is an understanding of how ubiquitous is the
jet phenomena for X-ray binaries.
}
\label{fig1}
\end{figure}

\section{Forwards}

In Fig 7 I attempt to briefly summarise the state of play of research
into X-ray binary jets. A key point is the ubiquity of jets; it is
hoped that I have by now convinced the reader that it is at least {\em
possible} if not {\em likely} that jets are important for the majority
of X-ray binary systems. Taking this as established, I have outlined
the areas in which important research can be done via either 

\begin{itemize}
\item{An empirical, `energetics' approach, in which the exact way in
which the jet extracts energy and angular momentum from the accretion
flow is not key, but rather estimating {\em how much} of these
quantities are associated with the jet, is.}
\item{A `physics' approach, in which we really try to probe the
physics of what is happening -- ie. how are the particles
accelerated, how are the inflow and outflow physically coupled, what
knowledge can we extract about solid surfaces/event horizons
associated with the compact object, etc.}
\end{itemize}

It is fair, I think, to say that while the second approach must be the
ultimate aim, it is only via the broader approach in which the
significance (without the details) can be established, that the
research interest in the detailed physics can be established.

\subsection{Energetic and dynamical significance of jets}

\subsubsection{Spectral extent and power of synchrotron component}

It was already noted that transient ejection events could require a
very large rate of power injection, and may be energetically
significant during the outburst phase of transients (e.g. Mirabel \&
Rodr\'\i guez 1994).  While Kaiser, Sunyaev \& Spruit (2000) argued
that the required energy input could be spread over a longer
timescale, at least in one case, the repeated oscillation/ejection
events in GRS 1915+105, this cannot be occurring (Fender \& Pooley
2000).

In the persistent sources, notably the black holes in Low/Hard states,
there are strong arguments that the self-absorbed synchrotron spectrum
extends to at least the near-IR or optical bands (Fender 2001;
Brocksopp et al. 2001; Fender et al. 2001). Since the jets are likely
to be radiatively inefficient (as are AGN jets, see e.g. Celotti this
volume), then the total jet power may begin to approach or even exceed
the broadband X-ray luminosity, traditionally taken to be the best
measure of accretion rate. In the case of the Low/Hard state transient
XTE J1118+480 (which has been dynamically established to contain a
black hole -- McClintock et al. 2001),
the ratio of total jet power to X-ray luminosity, if the
synchrotron spectrum extends {\em only} to the near-infrared, is of
order $P_{\rm J} / L_X \sim 0.01 \eta^{-1}$ where $\eta$ is the
radiative efficiency of the jet (Fender et al. 2001). For both X-ray
binaries and AGN, it is likely that $\eta < 0.1$, giving $P_{\rm J}
\geq 0.1 L_X$ -- so even in a very conservative estimate, the jet
power is at least 10\% of the total accretion luminosity and cannot
seriously be ignored.

Even more intriguingly, if the self-absorbed synchrotron component
extends to the optical band or beyond, then comparison of the
broadband radio--optical--X-ray spectra of BHCs in hard states show
that the optically thin component should have a significant role to
play in the X-ray band. In fact, Markoff, Falcke \& Fender (2001) show
that for XTE J1118 the jet can even fit almost the entire broadband
spectrum, dominating ($>$ 90\%) the power output of the system in the
low/hard X-ray state (Fig 8). For the neutron stars the data are
sparser, but Fomalont et al. (2001a,b) argue that the jet power in Sco
X-1 is at least an order of magnitude greater than the observed X-ray
luminosity. Furthermore, sources such as SS 433 and LS 5039 (Paredes
et al. 2000) are very weak in the X-ray band by the standards of XRBs
in general, yet are powerful producers of jets and maybe even
$\gamma$-rays. Perhaps we should adjust our thinking to consider that
X-rays are not the only tracers of accretion power at large in the
universe.

\begin{figure}
\begin{center}
\includegraphics[width=13cm, angle=0]{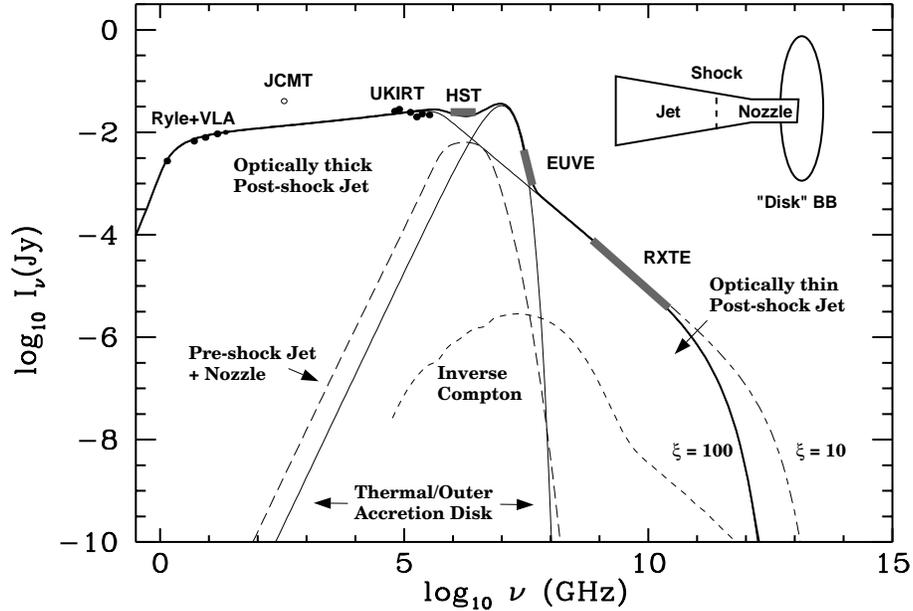}
\end{center}
\caption{
Broadband radio -- X-ray spectrum of the BHC XTE J1118+480 in the
low/hard X-ray state, fitted by a combination of a truncated accretion
disc and a jet. From Markoff et al. (2001).
}
\label{fig1}
\end{figure}

\subsubsection{Composition}

The question of whether or not XRB jets are in general comprised of a
normal baryonic (electrons + protons) plasma or of pairs (electrons +
positrons) is important both for our concept of the flow of mass in
accretion, and in estimating the energetics of the outflow. For
example, in the case of the repeated oscillations in GRS 1915+105 the
required power in the event that each oscillation is associated with
the relativistic bulk motion of a large number of protons is much
greater than if the plasma is simply electron:positron pairs (Fender
\& Pooley 2000).

Unfortunately, with the exception of SS 433 for which atomic emission
lines have been directly observed (e.g. Margon 1984), we only observe
synchrotron emission from electrons (and/or positrons) in the jets and
it is not straightforward to detect the presence of protons. One
ray of hope is that the use of circular polarisation measurements may
shed light on the composition of jets in both AGN and XRBs (e.g. 
Wardle 2001), although interpretations of data are not straightforward.
One XRB source, SS 433, has been detected in circular polarisation
(Fender et al. 2000a), but again the relation of this source to other
more `normal' XRBs is unclear, and more measurements are needed.

\subsubsection{Bulk velocity}

Essential to our understanding of the energetics of the outflow
($E_{\rm kinetic}\sim (\Gamma-1) \times E_{\rm internal}$), and also
for theoretical models of jet formation and their propagation through
the ISM, is the `terminal' Lorentz factor of the outflows in X-ray
binaries. Only SS 433 has a well-defined and only moderately
relativistic flow velocity. For other systems, even the well studied
ones such as GRS 1915+105, the `true' bulk Lorentz factor is such a
sensitive function of the assumed distance (Fender et al. 1999a) and
our interpretation of the data (Bodo \& Ghisellini 1995) that we
cannot really be certain at all of its value.

One interesting consideration is that the similarity of the
correlation between X-ray and radio fluxes in the Low/Hard state
sources Cyg X-1, GX 339-4 and others (Brocksopp et al. 1999; Corbel et
al. 2000; Fender 2001) may naively imply that one component cannot be
strongly beamed compared to the other, or the relation would be very
different from source to source, depending on inclination even if they
all had exactly the same velocity. Therefore it seems likely that the
bulk Lorentz factor is not exceptionally high, probably $<10$,
although a quantitative investigation is required to tell if the
available small-number statistics really are that constraining.  


\subsubsection{Jet rotation -- angular momentum transport}

While there is much progress in understanding the energetic
significance of jets from X-ray binaries, their influence (if any) on
the extraction of angular momentum from the accretion flow (a
necessary but poorly-understood process) remains unclear. Observations
of rotating jets from X-ray binaries, especially if the amount of
angular momentum could be quantitatively estimated, would be of great
significance.

\subsubsection{Interactions with the ISM}

Observations of interactions between XRB jets and the ISM are much
less common than observations of interactions between AGN jets and the
IGM; as a result we are left with one less diagnostic of the
energetics of the outflow -- in essence, the endpoint of the bulk of
XRB jet power (in the form of kinetic energy) is unknown. In a few
cases, interactions with the ISM have been observed -- the BHCs 1E
1740.7-2942 and GRS 1758-258 in the galactic centre show AGN-like
radio lobes (e.g. Mirabel 1994); SS 433 is clearly interacting with
the surrounding radio nebula W50 (e.g. Dubner et al. 1998); Cir X-1 is
surrounded by a radio nebula which seems to be powered by its radio
jets (Stewart et al. 1993; Fender et al. 1998), and there are a few
more examples. 

Besides being additional clues as to the total power of jets, other
intriguing possibilities exist which could be investigated by means of
the jet-ISM interaction. For example, are XRB jets a source of cosmic
rays ? do they induce star formation ? 

\section{Physical processes}

\subsection{Jet formation}

Detailed numerical modelling of relativistic jets currently favour
magnetohydrodynamic (MHD) models (e.g. Meier, Koide \& Uchida 2001 and
references therein). Can we test these models with observations of
X-ray binaries ? perhaps in some ways we can -- for example Meier
(2001) takes the empirically derived association between hard X-ray
states and radio emission in XRBs as some of the strongest
observational evidence for MHD jet formation in geometrically thick
accretion flows threaded by poloidal field lines.

\subsection{Comparison of neutron stars and black holes}

Are there any observed differences between the accretion:outflow
coupling in BHCs and NS systems ? There may be some hints -- firstly,
in Fender \& Hendry (2000) it was established that the persistent BHCs
in the Low/Hard state had approximately the same radio luminosity as
the Z sources on the HB. Since the Z sources are significantly more
luminous X-ray sources than the Low/Hard state BHCs, this already
implied that there was some difference. In Fender \& Kuulkers (2001)
the ratio of peak radio to X-ray flux was compared for all reported
(quasi-)simultaneous observations of X-ray transients. This ratio, or
`radio loudness' was found to be significantly higher for the
BHCs. There are two obvious possible causes of this effect -- either
BHCs are more efficient at producing jets
(extraction of energy from their deeper gravitational potentials?), or
maybe the BHCs are underluminous at X-ray wavelengths, even in
outburst, due to radiatively inefficient flows and their lack of a
solid surface. The former explanation implies that we are probing to
within the last few gravitational radii around the compact object; the
latter may be considered evidence for black hole event horizons.

\subsection{Particle acceleration and a comparison to AGN and GRBs}

Part of the theme of this volume is a comparison between the physics
of relativistic outflows from X-ray binaries, AGN and GRBs. I shall
briefly address both of these areas here.

\subsubsection{Particle acceleration}

Observation of the spectral index of optically thin synchrotron
sources allows a direct probe of the underlying electron population.
The synchrotron emission is produced by a (probably) nonthermal
(power-law) distribution of relativistic (Lorentz factors possibly to
$\geq 1000$) electrons spiralling in a magnetic field. Typically this
results in an observed power-law emission component at frequencies for
which self-absorption is not important (ie. `optically thin'), for
which we can define the spectral index $\alpha = \Delta \log S_{\nu} /
\Delta \log \nu$, i.e. the observed flux density $S_{\nu} \propto
\nu^{\alpha}$ (warning : many works, especially older papers on AGN,
use the reverse definition, ie. $S_{\nu} \propto \nu^{-\alpha}$).  If
the power-law distribution of electrons is described as $N(E)dE = N_0
E^{-p}dE$ (where $N(E)dE$ is the number of electrons with energy in
the range $E$ to $E+dE$, and $N_0$ is a constant), then the observed
spectral index $\alpha = (1-p)/2$ (for a plasma in which adiabatic
expansion losses dominate).  Thus measurement of the optically thin
spectral index directly provides information on the distribution of
relativistic electrons; typically $-1 \leq \alpha_{\rm opt. thin} \leq
-0.5$, implying $2 \leq p \leq 3$.

These values are broadly consistent with those predicted for
acceleration of the particles at a shock (e.g. Blandford \& Eichler
1987) and are comparable to those observed in AGN. For now it seems
reasonable to accept shock-accelerated power-law distributions of
electrons as the origin of the observed synchrotron emission.

\subsubsection{XRBs as mini-AGN: `Microquasars'}

The term `microquasar' is evocative and has been powerful in
attracting public and scientific interest to the field of jets from
X-ray binaries.  It cannot be denied that there is some accuracy in it
as a scientific expression, since in both AGN and X-ray binaries it
seems that an accretion flow around a black hole (or neutron star in
the case of the XRBs) results in the production of a powerful
collimated outflow. There are obvious differences too, such as the
supply of matter for accretion (or `Fuel Tank') which is clearly
different in the two cases, but perhaps, since the exotic physics
takes place relatively close to the black hole, where the material has
presumably `forgotten' where it came from, this is not important.  For
example, Falcke \& Biermann (1996) discuss the applicability of
scaling down AGN jet models to X-ray binaries.  It has often been
noted that since accretion timescales might be expected to scale with
mass of the accretor then processes which could never be observed from
an AGN in a single human lifetime may be observed many times over by
the same individual from a microquasar (e.g. Sams, Eckart \& Sunyaev
1996; Mirabel \& Rodr\'\i guez 1999).

Has the study of XRBs shed any light yet on the physics of AGN ? this
is less clear, but the prospects are good.  Certainly application of
many principles developed for AGN has been extremely useful in helping
us to understand XRB jets without having to reinvent the wheel, and it
would be nice to reciprocate.  An example may be the clear relation
between modes of accretion, or `states', and the presence of radio
jets in the BHC XRBs -- is this related to the radio-loud:radio-quiet
dichotomy in AGN ?

\subsubsection{XRBs and GRBs}

GRBs and their afterglows are now widely believed to be associated
with highly relativistic outflows (e.g. Sari, Piran \& Halpern 1999;
see also papers by Sari and Galama in these proceedings). Since they
appear to be small-scale objects (compared to AGN) it is therefore
natural to look for any connections with the jets of XRBs.
Pugliese, Falcke \& Biermann (1999) specifically discuss the 
possibility of a GRB from SS 433, and Portegies Zwart, Lee \& Lee
(1999) discuss the possibility of GRBs arising from a precessing
jet in a `gamma-ray binary'. For now the jury is out on the relevance
of such comparisons and models. In at least one respect, the bulk
Lorentz factor (invoked to be 100 or even more in GRBs) there seems
to be a significant distinction between the physics of the outflow
in the two types of object.

\section{Conclusions}

To conclude, the study of jets from X-ray binaries is providing much
exciting data on the physics of the coupling of accretion and outlflow
around both neutron stars and black holes. The latter in particular
provide hope that we may be able to learn about the physics of black
hole accretion in AGN, the powerhouses of the Universe, by studying
XRBs.  It is this author's feeling that jets will turn out to be a
fairly ubiquitous characteristic of the accretion process in XRBs. In
order to establish this, more observations of radio emission from the
NS Atoll sources, the single largest class of XRB, are required.

\section*{Acknowledgements}

I am very happy to acknowledge stimulating conversations with many of
my colleagues at the SURF2000 workshop in Mykonos. I would also like
to thank Sera Markoff for a careful reading of a draft of this
manuscript, and Qingzhong Liu for help with XRB population numbers.

%

\end{document}